# S-type Negative Differential Resistance in Semiconducting Transition-Metal Dichalcogenides

*Miao Wang[1][†], Chengyu Wang[1][†], Chenchen Wu[1], Qiao Li[1], Chen Pan[1], Cong Wang[1], Shi-Jun Liang[1,2]\*  and Feng Miao[2,1]\**

[1]National Laboratory of Solid State Microstructures, School of Physics, Collaborative Innovation Center of Advanced Microstructures, Nanjing University, Nanjing 210093, China.

[2]Shenzhen Research Insistute of Nanjing University, Shenzhen, 518057, China.

E-mail: miao@nju.edu.cn; sjliang@nju.edu.cn

† These authors contributed equally to this work.

Current-controlled (also known as "S-type") negative differential resistance (NDR) is of crucial importance to many emerging applications including neuromorphic computing and high-density memristors integration. However, the experimental realization of S-type NDR based on conventional mechanisms poses demanding requirements on materials, which greatly limits their potential applications. Here, we experimentally identify that semiconducting transition-metal dichalcogenides (TMDs) can host a bipolar S-type NDR devices. Theoretical simulations indicate that the origin of the NDR in these devices arises from a thermal feedback mechanism. Furthermore, we demonstrate the potential applications of TMDs based S-type NDR device in signal processing and neuromorphic electronics.

## 1. Introduction

S-type negative differential resistance (NDR) describes that an increase in current results in a reduction of voltage. Devices based on S-type NDR have been widely used in electronic circuits[1] including oscillators and amplifiers. In additional to the applications in existing technologies, S-type NDR also shows promising applications in emerging technologies. For instance, selector based on S-type NDR can limit the leakage current in memristor cross-bar arrays due to highly non-linear I-V characteristics[2-4]. S-type NDR generated neuron spikes allow for realization of neuromorphic electronic circuits[5, 6]. Most of reported S-type NDR result from distinct properties of materials, such as electronic instabilities[7, 8], interband tunneling[9, 10], threshold switching[11-13], insulator-metal transitions in metal oxides[14-25]. However, these mechanisms cannot be shared by most materials, restricting their wider applications. On the other hand, a recently proposed thermal feedback mechanism[4, 26, 27], in which only temperature dependence of material conductivity and Joule self-heating effect at large currents are involved, is a very promising approach for generating stable S-type NDR in most materials. Unfortunately, its experimental observation has been limited to the $NbO_x$ system so far[4, 26]. Therefore, it is quite desirable to expand the material matrix hosting the stable S-type NDR based on this new and simple mechanism. Here, we report the first experimentally identification of the family of semiconducting transition-metal dichalcogenides (TMDs) that can host the bipolar S-type NDR in a simple sandwich structure (electrode/TMD/electrode) device at room temperature. The observed NDR can be well described by the thermal feedback mechanism. Utilizing the semiconducting TMD-based NDR devices and several basic circuit elements, we develop a signal processing circuit capable of amplifying and reversing as well as multiplying AC signals, and a neuron circuit generating self-oscillation and neuron spikes.

## 2. Results and Discussions

The structure of the NDR devices is highly symmetric. We first deposited Au bottom electrodes (40 nm thick, 3 μm wide) on a 300-nm thick $SiO_2$ wafer through the standard Electron-beam lithography (EBL) and E-beam deposition processes. Mechanically exfoliated semiconducting TMD membrane (20~40 nm thick) was then transferred onto the Au bottom electrodes. An Au top electrode (40 nm thick, 3 μm wide) was finally deposited onto the TMD membrane to align with the bottom electrode. Figure. 1(a) shows an optical image of one typical NDR device. The $Au/MoS_2/Au$ device structure is shown in the inset of Figure. 1(b).

The repeatable I-V curve of $Au/MoS_2/Au$ devices was obtained by sweeping the applied current, with results shown in Figure. 1(b). The voltage increases with the applied current when the current is below ~2.5 mA. At higher current (>~2 mA), the voltage becomes negative correlation with the applied current. The I-V curve has no obvious hysteresis when sweeping current between 10 mA and zero. Note that $Au/WS_2/Au$ and $Au/WSe_2/Au$ devices also exhibit similar stable NDR (see Figure. S1(a) and Figure. S1(b) in the Supplementary Information), indicating that the observed NDR is an universal phenomenon in the semiconducting TMD materials. Furthermore, we used graphite as electrode to fabricate graphite/$MoS_2$/graphite (G/$MoS_2$/G) devices (see the experimental method in the Supplementary Information). Interestingly, we can still observe the stable NDR in such devices (see Figure. S1(c) in the Supplementary Information), ruling out the possibility that the electrode materials determine the presence of NDR. Once the applied current reaches a certain threshold value, all the devices would exhibit NDR and persist until the dielectric layers ($MoS_2$, $WS_2$, and $WSe_2$) are broken down under the large electric field and high temperature generated by the applied voltage and current.

Previous works have shown that the NDR in the metal/insulator/metal vertical structures can be attributed to the high current induced metal-insulator transition (MIT) of the insulator layers[14-25]. For this mechanism, NDR does not occur when the ambient temperature is higher than the transition temperature. However, the NDR in the Au/$MoS_2$/Au devices only disappears at the temperature of above 205 °C (symbols in Figure. 2), which is insufficiently high to induce phase change in $MoS_2$. Besides, the MIT-based mechanism usually gives rise to a significant hysteresis in the I-V curve, which is completely different from our experimental results. The NDR observed in our devices can be attributed to the thermal feedback mechanism caused by the self-Joule heating effect. As the current flowing through semiconducting TMD layer increases, the resulting Joule heating rises the temperature of the TMD layer. As a result, the resistance of TMD layer is reduced, which produces further increases in temperature of TMD layer. At a critical current, this positive feedback effect gives rise to the NDR.

Here, we assume that the temperature $T$ of the device is uniform and its variation with time can be described by[4]

$$C_{th}\frac{dT}{dt} = \frac{T_0 - T}{R_{th}} + IV. \qquad (1)$$

where $T_0$ is the ambient temperature, $C_{th}$ and $R_{th}$ is the effective thermal capacitance and resistance of the device, respectively.

We assume that the thermal resistance $R_{th}$ slowly varies with the temperature $T$ at the static limit. The temperature can be written as $T = T_0 + R_{th}IV$. For 2H phase TMDs, previous work[28] has shown that the conduction mechanism follows a thermal activation process (*i.e.* $R \propto e^{\frac{E}{k_B T}}$) along the direction perpendicular to the TMDs plane, indicating the dominant role of polaron conduction. Besides, the temperature dependence of carrier mobility[28] in the TMD along the same direction is determined as $\mu \propto T^{-n}$. Considering these results, the electric resistance $R$ along the direction perpendicular to the TMDs plane can be written as a function of temperature[29] $R = \beta T^n e^{\frac{E}{k_B T}}$, where

$\beta$ is a constant factor, $k_B$ is the Boltzmann's constant, $E$ is the corresponding activation energy. According to Ohm's law,

$$I = \frac{V}{\beta(T_0+R_{th}IV)^n} \cdot e^{-\frac{E}{k_B(T_0+R_{th}IV)}}. \tag{2}$$

The I-V curves at various ambient temperatures is in good agreement with the simulations results (see solid lines in Figure. 2) based on Eq. (2), with the fitting parameters as determined: $\beta = 0.055$, $R_{th} = 1.7 \times 10^4$ K/W, $E = 0.4$ eV, $n = 1.5$. The fitted thermal activation energy and the exponent $n$ are consistent with previous experiment[28]. It has been pointed that exponent n can be used to distinguish the adiabatic and non-adiabatic behavior of the polaron. $n = 1.5$ is an indication that conduction of small polaron is the dominant mechanism along the direction perpendicular to the TMDs plane. This is similar to that in the transition metal oxides TiO$_2$[30]. Note that we don't consider the effect of temperature on $R_{th}$ in the Eq. (2). Previous work has indicated that the temperature has a profound impact on $R_{th}$ at low temperature than high temperature region[31]. This could explain the slight difference between simulation and experimental results at lower temperature shown in Figure. 2. Besides, thickness variance of TMDs may affect the occurrence of NDR, as $R_{th}$ not only depends on the temperature but also the thickness. For a given temperature, the occurrence of NDR in the TMDs with different thickness requires different sets of current and voltage. Furthermore, we note that the onset of NDR under a given ambient temperature occurs at different voltages and currents for all TMDs used in this work. The can be interpreted by experimental fact that distinct TMDs have different $R_{th}$ at the same ambient temperature [31]. These differences in sets of voltages and currents for onset of NDR may be eliminated by choosing different thickness of TMDs and/or ambient temperature. Therefore, the technological usefulness of the onset of NDR in the different TMDs may not be affected in the practical applications, such as signal processing and neuromorphic computing.

As mentioned above, NDR device is able to amplify and reverse the AC signals. When a NDR device is connected in series with a resistor, the AC voltage across the NDR device ($v$) is given by

$$v = \frac{dr}{dr+dR_L}V. \tag{3}$$

where $V$ is the AC input voltage, $dr$ and $dR_L$ is the differential resistance of the NDR device and the resistor, respectively. When $dr < 0$ and $|dr| > |dR_L|$, we have $v > V$ based on Eq. (3), indicative of the amplified AC input signal. When $dr < 0$ and $|dr| < |dR_L|$ hold, we have $v/V < 0$ according to Eq. (3), which indicates that the AC input signal is reversed. With these principles, we built a G/MoS$_2$/G NDR device based signal processing circuit (Figure. 3(a)). A constant voltage source $V_C$ is added in series to the NDR device to control the current flowing through the NDR device. When G/MoS$_2$/G device is operated in the NDR region by controlling $V_C$, we are able to amplify and reverse the input signals by changing the resistance ($R_L$) of the load resistor (Figure. 3(c) and 3(d)).

The NDR devices fabricated in this work has highly repeatable I-V characteristics, thus allow us to obtain high quality output signals. Compared to our NDR devices, the I-V curves of MIT-based NDR devices usually have large hysteresis, which would result in an undesirable distortion of the output signal in signal processing circuits (e.g. frequency multipliers). For the fabricated NDR device in this paper, we can control $V_C$ to make sure that the current through NDR device is kept at the transition point between the positive and the negative differential resistance. If we increase the positive voltage of input AC signal, the increased current through NDR device would make the device to operate at the NDR region. Thus, the corresponding input voltage signal would be reversed by satisfying the condition of $|dr_{NDR}| < |dR_L|$. Conversely, the negative voltage of input signal would reduce the current through NDR device and cause it to enter the region of positive differential resistance, where the corresponding input voltage signal would not be reversed. With the properties shown above, we can generate a non-distortion output signal with doubling the frequency of input signal, as shown in Figure. 3(b).

Apart from the applications in signal processing circuit, the excellent properties of two-dimensional materials, such as superior mechanical strength, thermal stability, flexibility and so on, may enable better performance of electronic devices in artificial neurons over the NDR devices based on traditional inorganic materials. To show the potential application, we built a self-oscillating circuit based on G/MoS$_2$/G NDR devices (highlighted by blue dashed box in Figure. 4(a)). The NDR device could cancel the positive resistance of the circuit, creating a lossless resonator. With a constant voltage source ($V_C$) and a constant input current (I), the circuit could generate a stable AC signal output (Figure. 4(b)), the frequency $f_0$ of which is determined by the resonant frequency of the circuit $f_0 = \frac{1}{2\pi R_{NDR} C}$. We then connected a diode (D, 1N4007) and two resistors ($R_1$ and $R_2$) to the output of the self-oscillating circuit in series. The diode could rectify the AC signal and generate refractory periods (plateau between peaks in Figure. 4(c)), indicating the circuit shown in Figure. 4(a) could be used to simulate the neuron spikes in nervous system (Figure. 4(c)).

## 3. Conclusion

In summary, we realize the first experimental identification of the family of the semiconducting transition metal dichalcogenides hosting the S-type NDR based on the vertical structure of electrode/TMD/electrode. Combining experimental results and theoretical simulations, we attribute the NDR phenomenon to the thermal feedback conduction mechanism induced by the Joule self-heating effect. Furthermore, we demonstrate that semiconducting TMD materials show promising applications in signal processing and generation of neuron spikes for neuromorphic computing based on the S-type NDR. Our results greatly expand the materials matrix for realizing S-type NDR devices and pave the way for the development of neuromorphic electronics.

**Supporting Information**

Supporting Information is available from the Wiley Online Library or from the author.

**Acknowledgements**

M. W. and C.Y. Wang contributed equally to this work. This work was supported in part by the National Key Basic Research Program of China (2015CB921600), Shenzhen Basic Research Program (JCYJ20170818110757746), National Natural Science Foundation of China (61625402, 61574076), Natural Science Foundation of Jiangsu Province (BK20180330, BK20150055), Fundamental Research Funds for the Central Universities (020414380093, 020414380084), and Collaborative Innovation Center of Advanced Microstructures.

**Figure Captions**

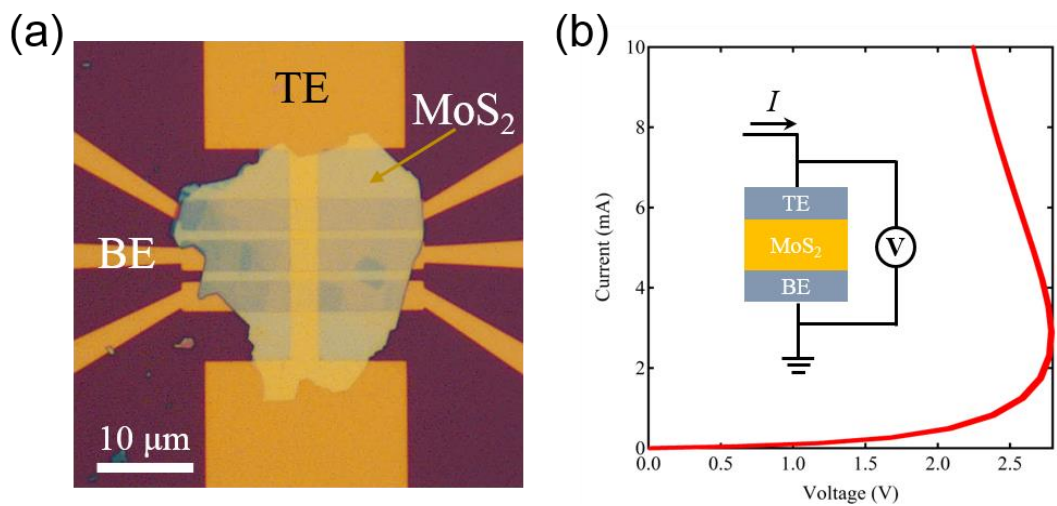

Figure. 1 Structure and characteristics of the NDR device based on $MoS_2$. (a) An optical image of the

NDR device. (b) Repeatable I–V curve of a typical Au/MoS$_2$/Au device, shows a S-type NDR. Inset: schematic drawing of the Au/MoS$_2$/Au device.

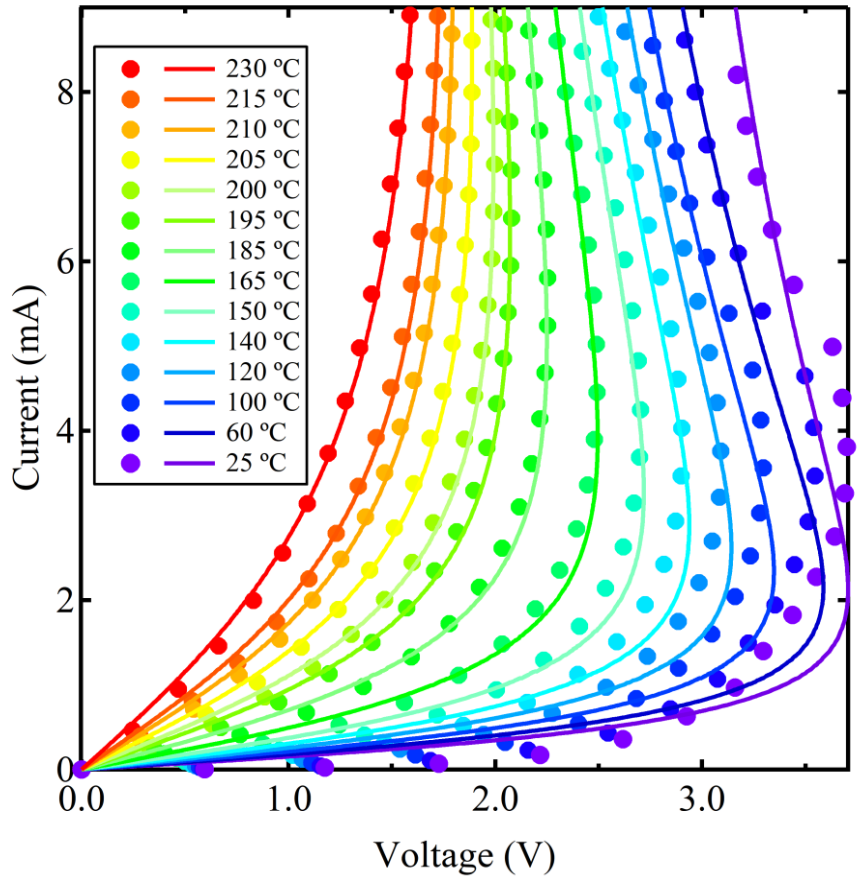

Figure. 2 Temperature dependence of the NDR device and theoretical simulations. Dots: Measured I-V curves of an Au/MoS$_2$/Au device at different temperatures. The NDR disappears at the temperatures of above 205 °C. Lines: Temperature dependent I-V characteristics for the device based on the Eq. (2).

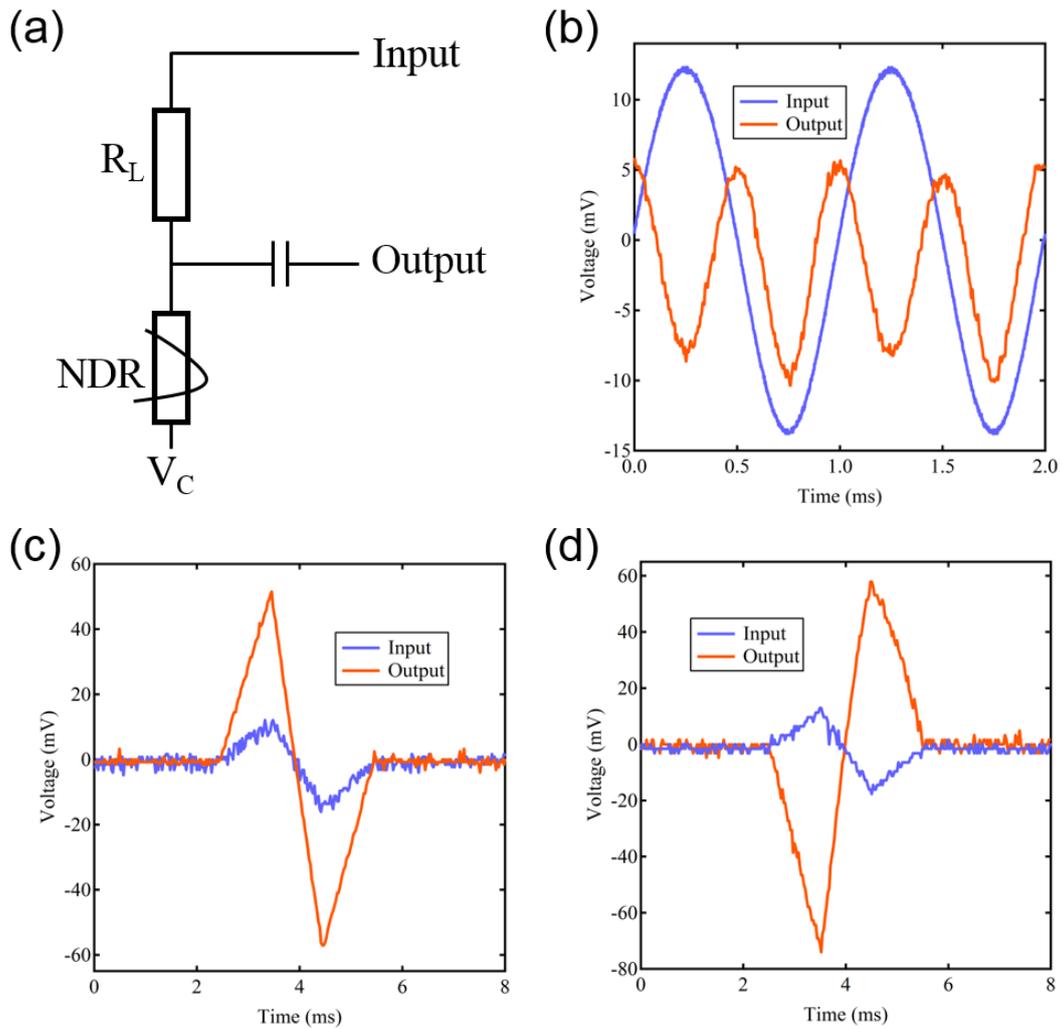

Figure. 3 Signal processing circuits. (a) Circuit layout based on G/MoS$_2$/G device for signal processing, where R$_L$ is a load resistor, V$_C$ is a constant voltage source, C is a capacitor (filtering DC voltage signals). (b) Frequency doubling of a sinusoidal voltage signal. (c) Amplification of a triangular voltage signal. (d) Inversion of a triangular voltage signal. The orange curve and blue curve corresponds to the input voltage signal and output voltage signal in (a), respectively.

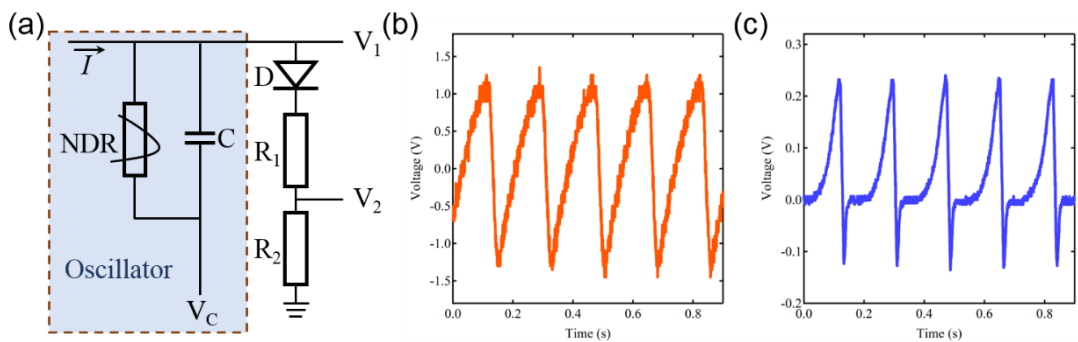

Figure. 4 Simulation of neuron spikes. (a) Circuit layout of simulating neuron spikes. The circuit consists of a self-oscillating circuit (highlighted by blue dashed box) based on G/MoS$_2$/G NDR device

and an output branch with a diode D (1N4007) in series with two ohmic resistors $R_1$ and $R_2$, which deliver the spikes from the oscillating voltage. Parameters used for the circuit: $V_C$ = -2.3 V, I = 8 mA, C = 100 μF, $R_1$ = 50 kΩ, $R_2$ = 10 kΩ. (b), (c) Recorded spike patterns for $V_1$ and $V_2$ in (a), respectively.